\documentclass{appolb}
\usepackage{graphicx}
\usepackage{bm}
\usepackage{amsmath}
\begin{document}
\title{Production of the Doubly Heavy Baryons, $B_c$ meson and the all-charm tetraquark at AFTER@LHC with double intrinsic heavy mechanism.%
}
\author{Sergey Koshkarev
\address{Institute of Physics, University of Tartu, Tartu 51010, Estonia}
\\
}
\maketitle
\begin{abstract}
In the paper we discuss contribution of the doubly intrinsic heavy mechanism into the production of  $B_c$ meson, the doubly heavy baryons and the all-charm tetraquark at a future fixed-target experiment at the LHC (AFTER@LHC). The production cross sections  and the mean values of Feynman-x  for the finale states are presented.
\end{abstract}
\PACS{PACS numbers come here}
 
 \section{Introduction}
\label{intro}

The low statistics NA3 experiment measurements of the double $J/\psi$ production~\cite{NA31982,NA31985} and the observation of the doubly charmed baryons by the SELEX collaboration~\cite{SELEX2002,Mattson,SELEX2005} cannot be described with the pertubative QCD.  In the Ref.~\cite{Vogt1995,Koshkarev} shown the importance of the the double intrinsic charm as the leading production mechanism. However this kind of physics is very limitedly accessible at the current collider experiments. The future high luminosity fixed-target experiment utilizing 7 TeV LHC beam (AFTER@LHC) can provide wide opportunities for a getting new data at the high Feynman-$x$ at a center-of-mass energy $\sqrt{s} = 115$ GeV~\cite{Brodsky2013,Lansberg2012,Lansberg2014,Rakotozafindrabe,Brodsky2015}.

The existence of a non-perturbative intrinsic heavy quark component in the nucleon is a rigorous prediction of QCD. Intrinsic charm and bottom quarks are in the wavefunction of a light hadron  --  from diagrams where the heavy quarks are multiply attached by gluons to the valence quarks~\cite{Brodsky1984,Franz}.  In this case, the frame-independent  light-front wavefunction of the light hadron has maximum probability when the Fock state  is minimally off-shell.  This occurs when all of  the constituents are at rest in the hadron rest frame  and thus have the same rapidity $y$ when the  hadron is boosted. Equal rapidity occurs when the light-front momentum  fractions $x={k^+\over P^+}$  of the Fock state constituents are proportional to their transverse mass: $ x_i \propto m_{T, i} = \sqrt { m^2_i + k^2_{T,i}}$; i.e. when the heavy constituents have the largest momentum fractions.  This feature underlies the Brodsky, Hoyer, Peterson, and Sakai (BHPS) model for the distribution of intrinsic heavy quarks~\cite{Brodsky1980,Brodsky1981}.

In the BHPS model the wavefunction of a hadron in QCD can be represented as a superposition of Fock state fluctuations, e.g. $| h \rangle \sim | h_l \rangle + | h_l g \rangle + | h_l Q \bar{Q} \rangle $... , where $Q=c,b$ and $ h_l$, as above, is light quark content. When the projectile interacts with the target, the coherence of the Fock components is broken and the fluctuation can hadronize. The intrinsic heavy flavor Fock components are generated by virtual interactions such as $gg \to Q \bar{Q}$ where gluon couple to two or more projectile valence quarks. The probability to produce such $ Q \bar{Q}$ fluctuations scales as $\alpha_s^2 (m_Q^2)/m_Q^2$ relative to leading-twist production.

Following \cite{Vogt1995,Brodsky1980,Brodsky1981} the general formula for the probability distribution of an n-particle intrinsic heavy flavor Fock state as a function of $x_i$  and transverse momentum $\vec{k}_{T,i}$ can be written as:
\begin{equation}
 \frac{dP_{iQ}}{\prod_{i=1}^n dx_i d^2 k_{T,i}} \propto  \alpha_s^4 (M_{Q \bar Q}) \frac{\delta \big(\sum_{i=1}^n \vec{k}_{T,i} \big)\delta \big( 1 - \sum_{i=1}^n x_i \big)}{\big( m_h^2 -  \sum_{i=1}^n m_{T,i}^2 / x_i \big)^2},
\end{equation}
where $m_h$ is mass of the initial hadron. The probability distribution for the production of two heavy quark pairs is
\begin{equation}
 \frac{dP_{iQ_1Q_2}}{\prod_{i=1}^n dx_i d^2 k_{T,i}} \propto  \alpha_s^4 (M_{Q_1 \bar{Q}_1}) \alpha_s^4 (M_{Q_2 \bar{Q}_2})  \frac{\delta \big(\sum_{i=1}^n \vec{k}_{T,i} \big)\delta \big( 1 - \sum_{i=1}^n x_i \big)}{\big( m_h^2 -  \sum_{i=1}^n m_{T,i}^2 / x_i \big)^2},
\end{equation}
 If we are interested in calculation of the $x$ distribution then we can simplify the formula with replacement $m_{T,i}$ with the effective mass $\hat{m}_i = \sqrt{m_i^2 + \langle k^2_{T,i} \rangle}$ and neglect the mass of the light quarks
\begin{equation}
 \frac{dP_{iQ_1Q_2}}{\prod_{i=1}^n dx_i} \propto  \alpha_s^4 (M_{Q_1 \bar{Q}_1}) \alpha_s^4 (M_{Q_2 \bar{Q}_2})  \frac{\delta \big( 1 - \sum_{i=1}^n x_i \big)}{\big(\sum_{i=1}^n \hat{m}_{T,i}^2 / x_i \big)^2}
\end{equation}
This model assumes that the vertex function in the intrinsic charm wavefunction is relatively slowly varying; the particle distributions are then controlled by the light-cone energy denominator and phase space. The Fock states can be materialized by a soft collision in the target which brings the state on shell. The distribution of produced open and hidden charm states will reflect the underlying shape of the Fock state wavefunction.

In this paper we shall investigate the doubly intrinsic heavy approach for the production of the doubly heavy states in the Feynman-$x$ region at a fixed-target experiment at LHC at $\sqrt{s} = 115$ GeV.

\section{The Double Heavy States Production Cross Section under the Quark-Hadron Duality}

\subsection{Production from $|uud c \bar c c \bar c \rangle$ Fock states}

In the frame of the quark-hadron duality the production cross section of the doubly heavy state can be estimated as the production of a heavy quark pair with small invariant mass between the heavy quark production threshold and the threshold of production open heavy flavor hadrons. In case of the doubly charmed baryons
\begin{equation}
m_c + m_c < M_{cc} < m_{th} = m_D + m_D ,
\end{equation}
where $m_c$ is c-quark mass, $m_D$ is D-meson mass. The production cross section of the doubly charmed baryons an be written as:
\begin{equation}
\sigma(\Xi_{cc}) \approx \frac{2}{3} \cdot f_{cc/p} \cdot \sigma_{icc},
\end{equation}
where the factor 2/3 comes from requirement of isolating color-antitriplet. The $cc$ pair has $3 \times 3 = 9$ color components, 3 color-antitriplet, and 6 color-sixtet. There is 1/3 probability for the color-antitriplet possibility. The $f_{cc/p}$ is the fraction of the $cc$ pairs in the duality interval produced by proton beam:
\begin{eqnarray}
f_{cc/p} \simeq \int_{4m_c^2}^{m_{th}^2} dM_{cc}^2 \frac{dP_{icc}}{dM_{cc}^2} \,\, \bigg/ \int_{4m_c^2}^{s} dM_{cc}^2 \frac{dP_{icc}}{dM_{cc}^2},
\end{eqnarray}
and $\sigma_{icc}$ is the double intrinsic charm cross section:
\begin{eqnarray}
\nonumber
\sigma_{icc} = \frac{P_{icc}}{P_{ic}} \cdot \sigma_{ic} \\
\label{eq:bbb}
 \sigma_{ic} = P_{ic} \cdot  \sigma^{in}_{pp} \, \frac{\mu^2}{4 \hat{m}_c^2} \, ,
\end{eqnarray}
$\mu^2 \approx 0.2$ GeV$^2$ denotes the soft interaction scale parameter; following the BHPS model we will assume $P_{ic} = 0.01$. In ref.~\cite{Vogt1995} it is found that for proton $P_{icc} \approx 20 \% \cdot P_{ic}$. The proton-proton inelastic scattering cross section $\sigma^{in}_{pp}$ is (for $\sqrt{s} \geq 100$ GeV)~\cite{Block}:
\begin{equation}
\sigma^{in}_{pp} = 62.59 \, \hat{s}^{-0.5} + 24.09 + 0.1604 \, \ln (\hat{s}) + 0.1433 \, \ln^2 (\hat{s}) \, \,\, \text{mb},
\end{equation}
where $\hat{s} = s/2m_p^2$.
At the AFTER@LHC energies, $\sqrt{s} = 115$ GeV, $\sigma^{in}_{pp}$ equals to 28.4 mb. The numerical value of the production cross section of the doubly charmed baryons will be:
\begin{eqnarray*}
\sigma(\Xi_{cc}) \approx 9.4 \times 10^4 \text{pb}.
\end{eqnarray*}

The $x_F$-distribution of $\Xi_{cc}$ baryons can be written as:
\begin{equation}
\frac{dP_{icc}(\Xi_{cc})}{dx} = \int \prod_{i=1}^n dx_i \frac{dP_{icc}}{dx_1 ... dx_n} \times \delta(x_{\Xi} - x_c -x_c).
\end{equation}
Figure 1 shows the normalized $x_F$-distribution. The mean $x_F$ value is 0.33. 
\begin{figure}[h]
\label{fig:shape}
\includegraphics[scale=0.7] {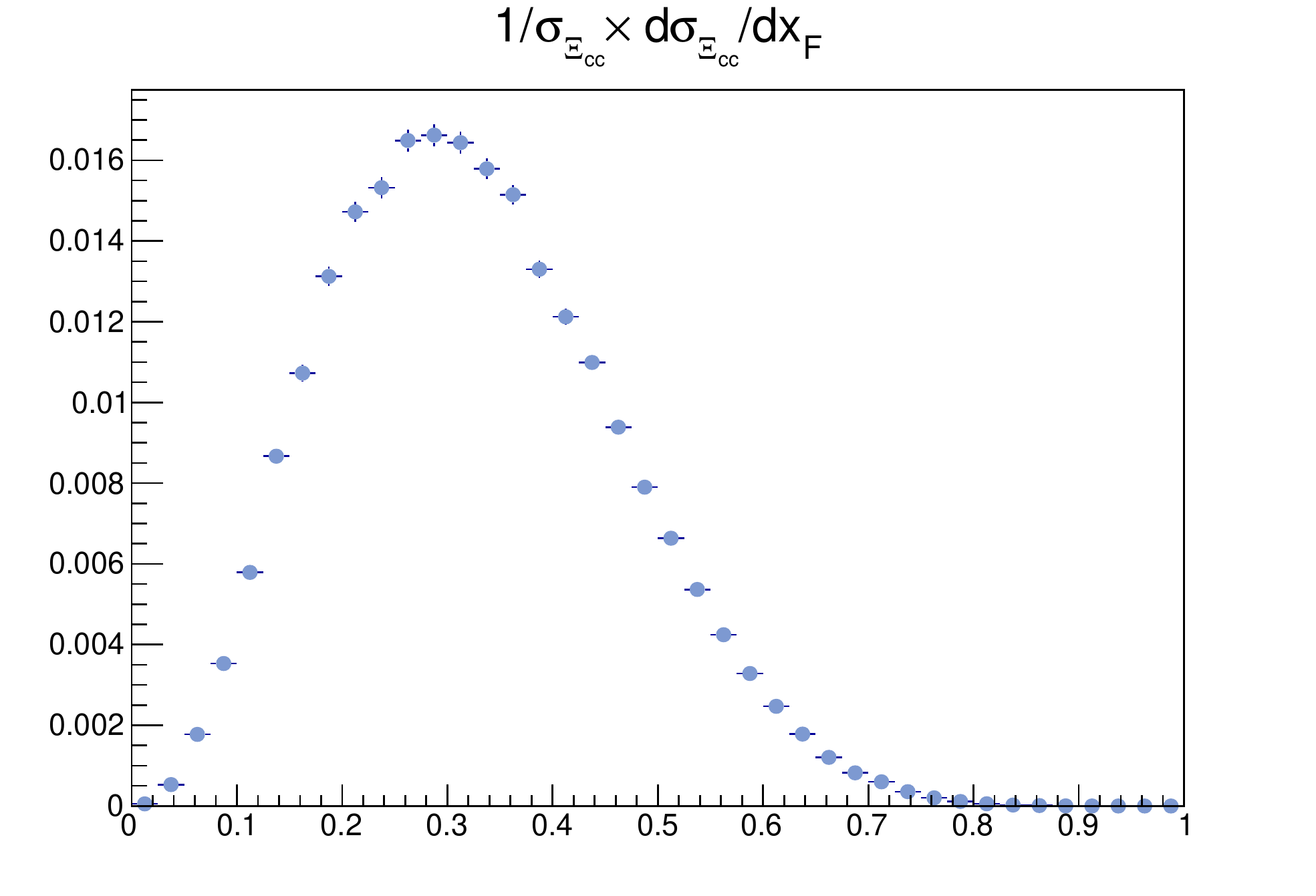}
\caption{The histogram represents calculation of the $x_F$-distribution of the doubly charmed baryons in the double intrinsic charm model.}
\end{figure}

It is interesting to estimate the production of the all-charm tetraquark $T_{4c}$ which can be naturally produced from the $|uud c \bar c c \bar c \rangle$ Fock state. The most recent prediction of the all-charm tetraquark scalar state mass is $m(T_{4c}) = 5.3 \pm (0.5)$ GeV~\cite{Heupel}. Then the duality interval will be
\begin{equation}
m_{J/\psi} + m_{J/\psi} < M_{c \bar c c \bar c} < m_{\Xi_{cc}^+} + m_{\Xi_{cc}^+},
\end{equation}
where $m_{J/\psi}$ and $m_{\Xi_{cc}^+} = 5.32$ GeV~\cite{SELEX2002} are masses of the $J/\psi$ meson and the lightest doubly charmed baryons respectively.
The production cross section will be 
\begin{equation*}
\sigma(T_c) \approx R \cdot f_{c \bar c c \bar c} \cdot \sigma_{icc} = R \times 10^4 \text{pb},
\end{equation*}
where the positive number $R < 1$. It used to be assumed 0.1. 
The Figure 2 shows the normalized $x_F$-distribution. The mean $x_F$ value is 0.67.
\begin{figure}[h]
\label{fig:shape2}
\includegraphics[scale=0.7] {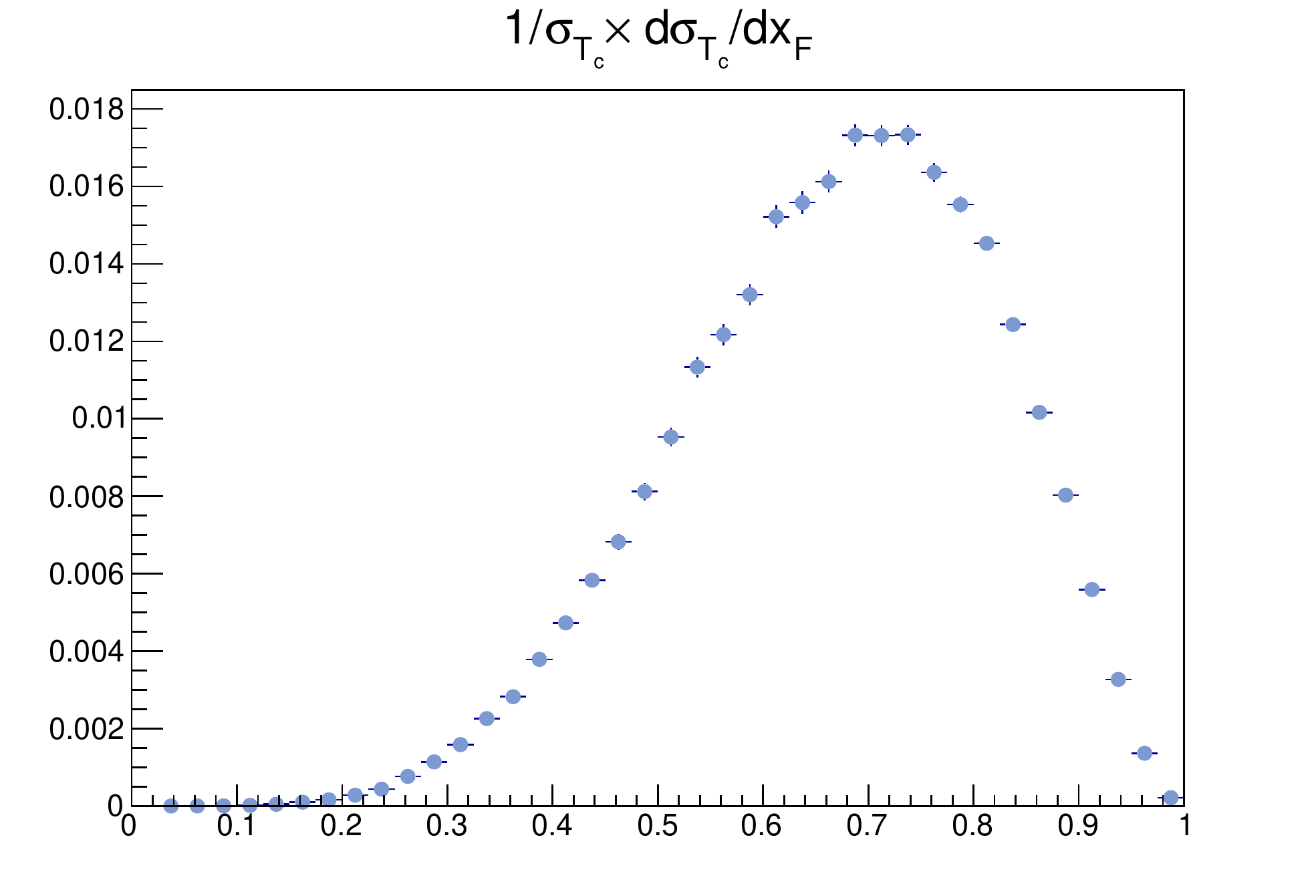}
\caption{The histogram represents calculation of the $x_F$-distribution of the all-charmed tetraquark ($T_{4c}$) in the double intrinsic charm model.}
\end{figure}

\subsection{Production from $|uud c \bar c b \bar b \rangle$ Fock states}

Following the method above we can calculate production of the $B_c$ meson and the beauty-charmed baryons. The duality interval for both states is
\begin{equation}
m_c + m_b < M < m_{th} = m_D + m_B,
\end{equation}
where $m_b$ is b-quark mass and $m_B$ is B-meson mass then obtain:
\begin{eqnarray}
\sigma(B_c) \approx f_{bc/p} \cdot \sigma_{icb} \\
\sigma(\Xi_{bc}) \approx \frac{2}{3} \cdot f_{\bar b c/p} \cdot \sigma_{icb} \, ,
\end{eqnarray}
where $f_{bc/p}$ and $f_{\bar b c/p}$ are respectively the fraction of the $bc$ and $\bar b c$ pairs in the duality interval. Following~\cite{Vogt1995_1,Vogt1995_2} the cross section for $| u u d c \bar{c} b \bar{b} \rangle$ configuration in a proton:
\begin{eqnarray}
\nonumber
\label{eq:bbb}
\sigma_{icb} = \frac{P_{icb}}{P_{ib}} \cdot \sigma_{ib}  \\
\sigma_{ib} = P_{ic} \, \sigma^{in}_{pp} \frac{\mu^2}{4 \hat{m}^2_b} \Bigg( \frac{\hat{m_c}}{\hat{m_b}} \Bigg)^4
\Bigg( \frac{\alpha_s (\hat{m}_{b \bar b})}{\alpha_s (\hat{m}_{c \bar c})} \Bigg)^4 \, ,
\end{eqnarray}
Assuming that $P_{icb}/P_{ib} \approx P_{icc}/P_{ic}$, then obtain the numerical value of the production cross section:
\begin{eqnarray}
\sigma(B_c) \approx 3.9 \times 10^2 \text{pb} \\
\sigma(\Xi_{bc}) \approx 2.6 \times 10^2 \text{pb}.
\end{eqnarray}
The $x_F$-distribution is similar to that distribution of the doubly charmed baryons (see Fig. 1) but the mean value is 0.37.

\subsection{Production from $|uud b \bar b b \bar b \rangle$ Fock states}

It is easy to estimate the production of the doubly beauty baryons from the $|uud b \bar b b \bar b \rangle$ Fock states. The duality interval will be $m_b + m_b < M < m_{th} = m_{\Upsilon} + m_{\Upsilon}$. The production cross section of the doubly beauty baryons as above is
\begin{equation}
\sigma(\Xi_{bb}) \approx \frac{2}{3} \cdot f_{\bar b b/p} \cdot \sigma_{ibb},
\end{equation}
where following the BHPS model $\sigma_{ibb} = ( \hat{m}_c / \hat{m}_b )^2 \cdot \sigma_{icb}$. The numerical value of the production cross section of the doubly beauty baryon is 50 pb.

\section{Summary}

We present the production properties of $B_c$ meson, the doubly heavy baryons and the all-charm tetraquark $T_{4c}$ at a future fixed-target experiment AFTER@LHC. It is clear that not all heavy diquarks hadronize into proposed finale states therefore we need to interpret the presented cross sectiona as the upper limit. Also it is interesting to compare our predictions with the single intrinsic charm mechanism predictions given in Ref~\cite{GuChen} (see Table 1).

\begin{table}%[h!]
\centering
\caption{The table presents production cross section of the doubly heavy baryons at the AFTER@LHC. First column shows predictions in the double heavy intrinsic model and the second present predictions in the single intrinsic charm model}
\label{my-label}
\begin{tabular}{|| c c c ||}
 Particle type&this work &Ref.$~\cite{GuChen}$ \\
 $\Xi_{cc}$&$9.4 \times 10^4~\text{pb}$&$4 \times 10^3~\text{pb}$  \\
 $\Xi_{bc}$&$2.6 \times 10^2~\text{pb}$&$8.95~\text{pb}$  \\
 $\Xi_{bb}$& 50~\text{pb}&$3.1 \times 10^{-2}~\text{pb}$ 
\end{tabular}
\end{table}
It is easy to see that the doubly intrinsic heavy mechanism become the leading production mechanism at high Feynman-$x$ (see also discussion in Ref.~\cite{Groote}).


\begin{thebibliography}{00}

\bibitem{NA31982}
	J.~Badier {\sl et al.}, Phys. Lett. B{\bf114}, 457 (1982).

\bibitem{NA31985}
	J.~Badier {\sl et al.}, Phys. Lett. B{\bf158}, 85 (1985).

\bibitem{SELEX2002}
	M.~Mattson {\sl et al.} (SELEX~Collaboration), Phys. Rev. Lett. 89,  112001 (2002),~ArXiv:hep-ex/0208014.

\bibitem{Mattson}
	M.~Mattson, Ph.D. thesis, Carnegie Mellon University, 2002.

\bibitem{SELEX2005}
	A.~Ocherashvili {\sl et al.} (SELEX~Collaboration), Phys. Lett. B{\bf628}, 12-24 (2005),~ArXiv:hep-ex/0406033.
	
\bibitem{Vogt1995}
	R. Vogt and S. Brodsky, Phys. Lett. B349, 569-575 (1995), ArXiv:hep-ph/9503206.
	
\bibitem{Koshkarev}
	S.~Koshkarev and V.~Anikeev, ArXiv:1605.03070.

\bibitem{Brodsky2013}
	S. Brodsky, F. Fleuret, C. Hadjidakis, J. Lansberg, Phys. Rep. 522, 239-255 (2013).
	
\bibitem{Lansberg2012}
	J. Lansberg, S. Brodsky, F. Fleuret, and C. Hadjidakis, Few-Body Systems, vol. 53, no. 1-2, pp. 11-25 (2012).
	
\bibitem{Lansberg2014}
	J. Lansberg {\sl et al.}, EPJ Web of Conferences, vol. 66, Article ID 11023 (2014).
	
\bibitem{Rakotozafindrabe}
	A. Rakotozafindrabe, M. Anselmino, R. Arnaldi et al., Proceedings of the 21st International Workshop on Deep-Inelastic Scattering and Related Subject (DIS \textsc{\char13}13), p. 250 (2013).

\bibitem{Brodsky2015}
	S. J. Brodsky {\sl et al.}, Advances in High Energy Physics, Volume 2015, Article ID 231547 (2015).
	
\bibitem{Brodsky1984}
	S.~Brodsky {\sl et al.}, CNUM: C84-06-23 (1984).

\bibitem{Franz}
	M.~Franz {\sl et al.}, Phys. Rev. D{\bf62}, 074024 (2000).
	
\bibitem{Brodsky1980}
	S.~Brodsky {\sl et al.}, Phys. Lett. B{\bf93}, 451 (1980).

\bibitem{Brodsky1981}
	S.~Brodsky {\sl et al.}, Phys. Rev. D{\bf23}, 2745 (1981).
	
\bibitem{Block}
	M.~Block and F.~Halzen, Phys. Rev. D{\bf86}, 014006 (2012).
	
\bibitem{Heupel}
	W.~Heupel, G.~Eichmann, C.~Fischer, Phys. Lett. B{\bf718}, 545-549 (2012).

\bibitem{Vogt1995_1}
	R.~Vogt and S.~Brodsky, Nucl. Phys. B438, 159-188 (1995).

\bibitem{Vogt1995_2}
	R.~Vogt, Nucl. Phys. B446, 159-188 (1995).


\bibitem{GuChen}
	Gu~Chen {\sl et al.}, Phys. Rev. D 89, 074020 (2014), ArXiv:1401.6269.

\bibitem{Groote}
	S.~Koshkarev and S.~Groote, Nucl. Phys. B915, 384-391 (2017), ArXiv:1611.08149.


\end{thebibliography}
\end{document}